\documentclass[12pt]{article}
\usepackage{epsfig}
\def\be{\begin{equation}}
\def\ee{\end{equation}}
\def\bea{\begin{eqnarray}}
\def\eea{\end{eqnarray}}
\usepackage{graphicx}
\usepackage{color}
\usepackage{slashed}
\usepackage{amsfonts}

\catcode`\@=11
\def\lsim{\mathrel{\mathpalette\@versim<}}
\def\gsim{\mathrel{\mathpalette\@versim>}}
\def\@versim#1#2{\vcenter{\offinterlineskip
\ialign{$\m@th#1\hfil##\hfil$\crcr#2\crcr\sim\crcr } }}
\catcode`\@=12

\catcode`@=12
\begin{document}
\thispagestyle{empty}
\begin{flushright}
UCRHEP-T558\\
June 2016\
\end{flushright}
\vspace{0.6in}
\begin{center}
{\LARGE \bf Verifiable Associated Processes from\\ Radiative Lepton Masses 
with Dark Matter\\}
\vspace{1.5in}
{\bf Sean Fraser, Ernest Ma, and Mohammadreza Zakeri\\}
\vspace{0.2in}
{\sl Department of Physics and Astronomy, University of California,\\
Riverside, California 92521, USA\\}
\end{center}
\vspace{1.5in}
\begin{abstract}\
If leptons do not couple directly to the one Higgs doublet of the standard 
model of particle interactions, they must still do so somehow indirectly 
to acquire mass, as proposed recently in several models where it happens 
in one loop through dark matter.  We analyze the important consequences 
of this scenario in a specific model, including Higgs decay, muon anomalous 
magnetic moment, $\mu \to e \gamma$, $\mu \to eee$, and the proposed dark 
sector.
\end{abstract}

\newpage
\baselineskip 24pt

\section{Introduction}

The idea that lepton masses are induced in one loop has been around for 
a long time.  Recently it has been proposed~\cite{m14,m15,m15-1} that the 
particles in the loop are distinguished from ordinary matter by an unbroken 
symmetry so that the lightest neutral particle among them may be the dark 
matter of the Universe.  As an example, consider the specific proposal 
of Ref.~\cite{m15-1} for generating charged-lepton masses.
\begin{figure}[htb]
\vspace*{-3cm}
\hspace*{-3cm}
\includegraphics[scale=1.0]{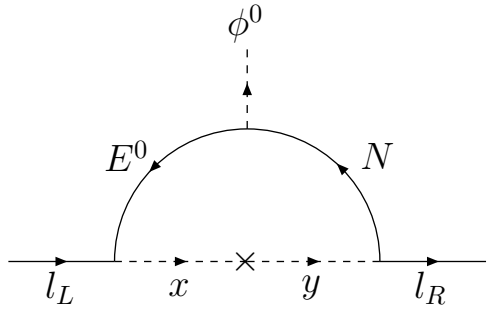}
\vspace*{-21.5cm}
\caption{One-loop generation of charged-lepton mass.}
\end{figure}
This model assumes the non-Abelian discrete symmetry $A_4$ under which the 
three families of leptons transform as
\begin{equation}
(\nu_i,l_i)_L \sim \underline{3}, ~~~ l_{iR} \sim \underline{1}, \underline{1}', 
\underline{1}''.
\end{equation}
With only the one Higgs doublet $(\phi^+,\phi^0)$ of the standard model (SM) 
transforming as $\underline{1}$, a tree-level lepton mass is forbidden. 
To obtain one-loop radiative lepton masses, the following new particles are 
added, all of which are odd under an unbroken dark $Z_2$ symmetry:
\begin{equation}
(E^0,E^-)_{L,R} \sim \underline{1}, ~~~ N_{L,R} \sim \underline{1}, ~~~ 
x^-_i \sim \underline{3}, ~~~ y^-_i \sim \underline{1}, \underline{1}', 
\underline{1}'',
\end{equation}
where $(E^0,E^-),N$ are fermions and $x^-,y^-$ are charged scalars.  
Note that in supersymmetry, there are also similar new particles, i.e. 
left and right charged sleptons and doublet Higgsinos.  The soft 
breaking of $A_4$ to $Z_3$ lepton triality~\cite{m10,cdmw11} is encoded in 
the scalar off-diagonal mass-squared $x_i y^*_j$ terms. 
In this paper we will study the phenomenological consequences of this 
proposal, including the deviation of the Higgs to charged-lepton decay from 
the SM, the muon anomalous magnetic moment, $\mu \to e \gamma$, 
$\mu \to eee$, as well as the structure of its dark sector.

\section{Radiative Lepton Masses}

The mass matrix linking $(\bar{N}_L, \bar{E}^0_L)$ to $(N_R, E^0_R)$ 
is given by
\begin{equation}
{\cal M}_{N,E} = \pmatrix{m_N & m_D \cr m_F & m_E},
\end{equation}
where $m_N,m_E$ are invariant mass terms, and $m_D,m_F$ come from the Higgs 
Yukawa terms $f_D \bar{N}_L E^0_R \bar{\phi}^0$, $f_F \bar{E}_L^0 N_R \phi^0$ 
with vacuum expectation value $\langle \phi^0 \rangle = v/\sqrt{2}$ .  
As a result, $N$ and $E^0$ mix to form two Dirac fermion eigenstates
\begin{equation}
n_{1(L,R)} = \cos \theta_{L,R} N_{L,R} - \sin \theta_{L,R} E^0_{L,R}, ~~~ 
n_{2(L,R)} = \sin \theta_{L,R} N_{L,R} + \cos \theta_{L,R} E^0_{L,R},
\end{equation}
of masses $m_{1,2}$, with mixing angles
\begin{eqnarray}
m_D m_E + m_F m_N &=& \sin \theta_L \cos \theta_L (m_1^2 - m_2^2), \\
m_D m_N + m_F m_E &=& \sin \theta_R \cos \theta_R (m_1^2 - m_2^2).
\end{eqnarray}
With the $A_4$ assignment of Eq.~(2), and the soft breaking to $Z_3$ of 
the term $x_i y_j^*$, i.e.
\begin{equation}
U_\omega \pmatrix{\mu_e^2 & 0 & 0 \cr 0 & \mu_\mu^2 & 0 \cr 0 & 0 & \mu_\tau^2} 
= {1 \over \sqrt{3}} \pmatrix{ 1 & 1 & 1 \cr 1 & \omega & \omega^2 
\cr 1 & \omega^2 & \omega} \pmatrix{\mu_e^2 & 0 & 0 \cr 0 & \mu_\mu^2 & 0 
\cr 0 & 0 & \mu_\tau^2},
\end{equation}
where $\omega = \exp (2 \pi i/3) = -1/2 + i \sqrt{3}/2$, and $U_\omega$ is 
the familiar~\cite{mr01} unitary matrix derivable from $A_4$,   
the charged-lepton mass matrix is given by
\begin{equation}
{\cal M}_l = U^\dagger_\omega \pmatrix{m_e & 0 & 0 \cr 0 & m_\mu & 0 \cr 
0 & 0 & m_\tau},
\end{equation}
with
\begin{equation}
m_e = -if' f_e \mu_e^2 \int {d^4 k \over (2 \pi)^4} {1 \over (k^2 - m_{1e}^2)
(k^2 - m_{2e}^2)} \left[ {m_1 \cos \theta_R \sin \theta_L \over k^2 - m_1^2} 
- {m_2 \cos \theta_L \sin \theta_R \over k^2 - m_2^2} \right],
\end{equation}
where $f'$ is the $E^0_L l_L x^*$ Yukawa coupling, $f_e$ is the $N_R e_R y^*_1$ 
Yukawa coupling, and $m_{1e,2e}$ are the mass 
eigenvalues of the $2 \times 2$ mass-squared matrix
\begin{equation}
{\cal M}^2_{xy_1} = \pmatrix{m_x^2 & \mu_e^2 \cr \mu_e^2 & m_{y_1}^2},
\end{equation}
with $\mu_e^2 = \sin \theta_e \cos \theta_e (m^2_{1e} - m^2_{2e})$, and 
similarly for $m_\mu$ and $m_\tau$.   It is clear that the residual $Z_3$ 
triality~\cite{m10,cdmw11} remains exact with $e,\mu,\tau \sim 1, \omega^2, 
\omega$, and the Higgs coupling matrix as well as the anomalous magnetic 
moment matrix are diagonal, as far as Fig.~1 is concerned.  In other words, 
flavor is not violated in Higgs decays and $\mu \to e \gamma$ is not 
mediated by the new particles of Eq.~(2).

\section{Anomalous Higgs Yukawa Couplings}

One immediate consequence of a radiative charged-lepton mass 
is that the Higgs Yukawa coupling $h \bar{l} l$ is no longer exactly $m_l/v$ 
as in the SM.  Its deviation is not suppressed by the usual one-loop 
factor of $16 \pi^2$ and may be large enough to be observable~\cite{fm14}. 
Moreover, this deviation is finite and calculable exactly in one loop. 
For discussion, compare our proposal to the usual consideration of the 
deviation of the Higgs coupling from $m_l/v$ from new physics in terms of 
higher-dimensional operators, i.e. 
\begin{equation}
-{\cal L} = f_l \bar{l}_L l_R \phi^0 \left( 1 + 
{\Phi^\dagger \Phi \over \Lambda^2} \right),
\end{equation}
where $\Lambda^2 >> v^2$.  This implies $m_l = (f_l v/\sqrt{2}) 
(1 + v^2/2\Lambda^2)$, whereas the Higgs coupling is $(f_l/\sqrt{2}) 
(1 + 3v^2/2\Lambda^2) \simeq (m_l/v) (1 + v^2/\Lambda^2)$.  However, 
this approach is only valid for $v^2 << \Lambda^2$, which guarantees 
the effect to be small. 
In the present case, if our result is interpreted as an expansion in 
powers of $v^2$, then it is a sum of infinite number of terms 
for both $m_l$ and the Higgs coupling, but each sum is finite. 
Their ratio is not necessarily small because some particles in the loop 
could be light, as shown below.

There are three contributions to the $h \bar{l} l$ coupling: (1) the 
Yukawa terms $(f_D/\sqrt{2}) h \bar{N}_L E_R^0$ and $(f_F/\sqrt{2}) 
h \bar{E}_L^0 N_R$, (2) the scalar trilinear $(\lambda_x v) h x^* x$ term, and 
(3) the scalar trilinear $(\lambda_y v) h y^* y$ term. 
In the following expressions, the couplings $f_{D,F}$ 
do not appear explicitly because
they have been expressed in terms of the
fermion masses $m_{1,2}$ and angles $\theta_{L,R}$. 
Consider $h \bar{\tau} \tau$.  The first contribution is given by
\begin{equation}\label{eq1}
f_{\tau}^{(1)}=\frac{f'f_{\tau}\sin2\theta_{\tau}}{32\pi^2v}\big[c_Rs_LT_1+s_Ls_RT_2+c_Lc_RT_3+c_Ls_RT_4\big],
\end{equation}
where $x_{ij}= (\frac{m_{i \tau}}{m_{j}})^2$,~ $s_{L,R}=\sin\theta_{L,R},~ c_{L,R}=\cos\theta_{L,R}$  and
\begin{eqnarray}\label{eq2}
F_N(x) &=&  {x(1+x)\ln x \over (1-x)^2}+{2 \over 1-x},~~~ H(x)= \frac{x}{x-1}\ln x\nonumber \\
T_1 &=& [2m_2s_Lc_Ls_Rc_R-m_1(s^2_Lc^2_R + c^2_Ls^2_R)][F_N(x_{11})-F_N(x_{21})],\nonumber\\
T_2 &=& m_2s_Lc_L(c^2_R-s^2_R)[H(x_{22})-H(x_{12})]-m_1s_Rc_R(c^2_L-s^2_L)[H(x_{21})-H(x_{11})],\nonumber\\
T_3 &=& m_1s_Lc_L(c^2_R-s^2_R)[H(x_{21})-H(x_{11})]-m_2s_Rc_R(c^2_L-s^2_L)[H(x_{22})-H(x_{12})],\nonumber\\
T_4 &=& [2m_1c_Lc_Rs_Ls_R-m_2(s_L^2c_R^2+c_L^2s_R^2)][F_N(x_{12})-F_N(x_{22})].
\end{eqnarray}
The second contribution is given by
\begin{equation}\label{eq3}
f_{\tau}^{(2)}=\frac{\lambda_xvf'f_{\tau}\sin2\theta_{\tau}s_Lc_L}{32\pi^2m_1m_2}\big[c^2_{\tau}T_1'+s^2_{\tau}T_2'\big],
\end{equation}
where
$c_{\tau} = \cos \theta_{\tau}$,
$s_{\tau} = \sin \theta_{\tau}$ and
\begin{eqnarray}\label{eq4}
F(x,y)&=&\frac{1}{x-y}\left[\frac{x}{x-1}\ln x-\frac{y}{y-1}\ln y \right]~~ x\neq y,~~F(x,x)=\frac{1}{x-1}-\frac{\ln x}{(x-1)^2},\nonumber\\
T_1'&=&m_2[F(x_{11},x_{11})-F(x_{11},x_{21})]-m_1[F(x_{12},x_{12})-F(x_{12},x_{22})],\nonumber \\
T_2'&=&m_2[F(x_{11},x_{21})-F(x_{21},x_{21})]-m_1[F(x_{12},x_{22})-F(x_{22},x_{22})].
\end{eqnarray}
The third contribution is given by
\begin{equation}\label{eq5}
f_{\tau}^{(3)}=\frac{\lambda_yvf'f_{\tau}\sin2\theta_{\tau}s_Lc_L}{32\pi^2m_1m_2}\big[s^2_{\tau}T_1'+c^2_{\tau}T_2'\big].
\end{equation}
Combining all three contributions and using Eq. (9) for the tau mass, 
the effective Higgs Yukawa coupling ${\tilde f}_{\tau}$ is given by
\begin{eqnarray}\label{eq6}
\frac{{\tilde f}_{\tau}v}{m_{\tau}} &=& \frac{[f_{\tau}^{(1)}+f_{\tau}^{(2)}+f_{\tau}^{(3)}]v}{m_{\tau}}\nonumber\\
&=& \frac{c_Rs_LT_1+s_Ls_RT_2+c_Lc_RT_3+c_Ls_RT_4+\frac{v^2s_Lc_L}{m_1m_2}[(\lambda_xc^2_{\tau}+\lambda_ys^2_{\tau})T'_1+(\lambda_xs^2_{\tau}+\lambda_yc^2_{\tau})T'_2]}{s_Lc_Rm_1[H(x_{21})-H(x_{11})]+s_Rc_Lm_2[H(x_{12})-H(x_{22})]}.\nonumber\\
\end{eqnarray}
\begin{figure}[htb]
\centering
\includegraphics[scale=.65]{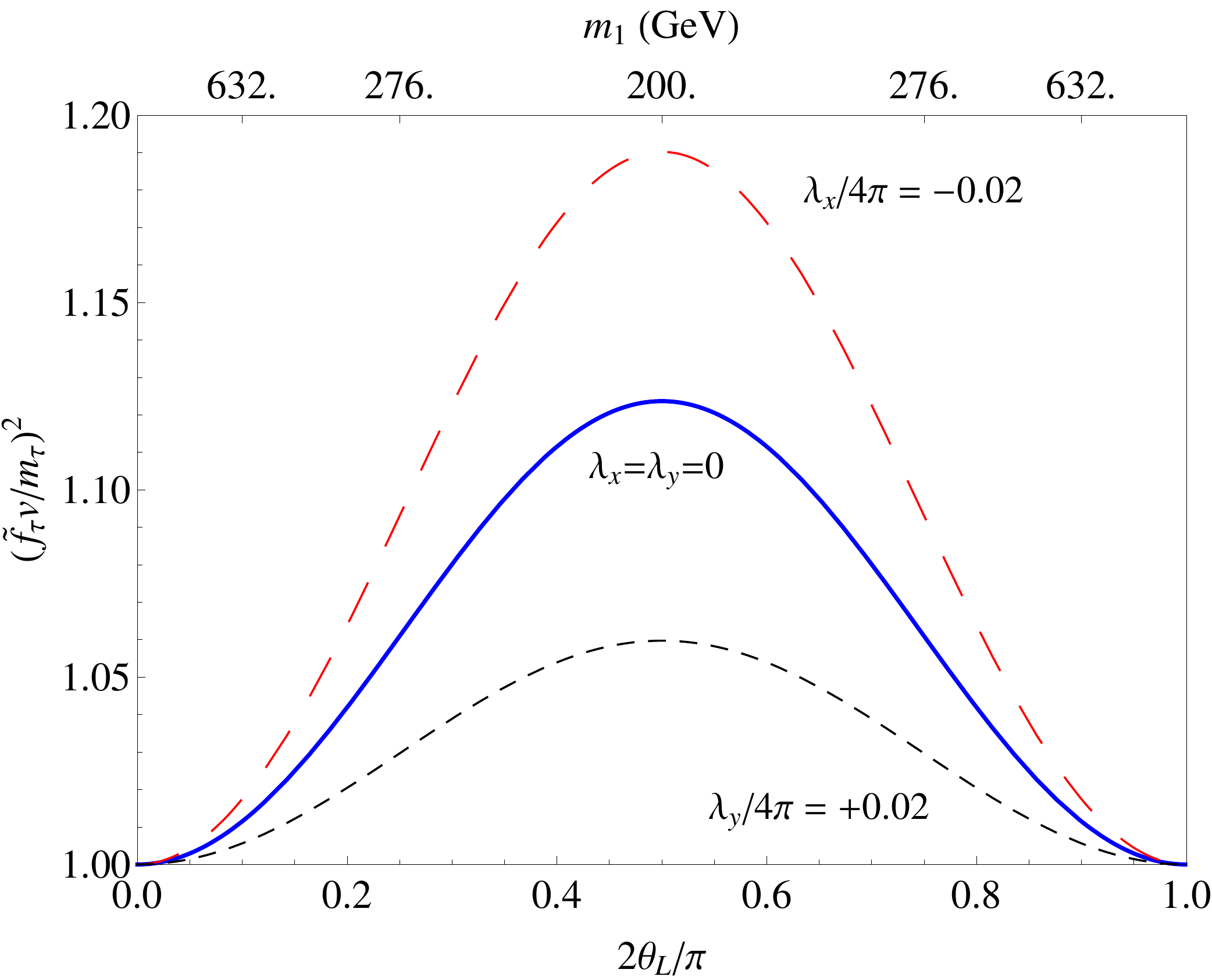}
\caption{The ratio $( {\tilde f}_{\tau} v/m_{\tau} )^2$
plotted against $\theta_L$ with various $\lambda_{x,y}$ 
for the case $\theta_L=\theta_R$.}
\end{figure}

To simplify the analysis, we focus on $\theta_L=\theta_R$, 
in which case $f_D=f_F$. 
We use the relation 
$ f_D  v / \sqrt{2} = s_L c_L(m_1 - m_2) = s_L c_L m_1(1 - m_2/m_1)$
from fermion mixing to define $m_1$ 
as a function of $\theta_L$ 
for a constant ratio $m_2/m_1=2.2$ and coupling
$f_D/\sqrt{4 \pi} = -0.19$.
In this parameterization, the combination $s_L c_L m_1$
remains constant, and also appears in the radiative mass formula
for each charged lepton. In addition, we use the value 
$f'/\sqrt{4 \pi} = -0.6$.
For the scalars in the tau sector, we choose fixed mass ratios 
$m_{1 \tau} / m_1 = 5.7$ and  
$m_{2 \tau} / m_1 = 1.1$. To satisfy the mass formula,
we verify that the product
$f_{\tau}\sin2\theta_{\tau}$ is not too large.
We have checked that the values used here also allow solutions for the 
muon and electron radiative masses. 
In Fig. 2 we plot the effective Yukawa coupling from Eq. (17)
as a function of $\theta_L$,
using the values $f_{\tau}/\sqrt{4 \pi}=-0.54$, $\theta_{\tau}=0.8$ for the 
$\lambda_{x,y}$ curves. We see that a significant deviation from the SM prediction 
is possible.

\section{Muon Anomalous Magnetic Moment}

Another important consequence of a radiative charged-lepton mass is that 
the same particles which generate $m_l$ also contribute to its anomalous 
magnetic moment.  This differs from the usual contribution of new physics, 
because there is again no $16 \pi^2$ suppression. There are three contributions to the anomalous magnetic moment. The main contribution is given by
\begin{equation}\label{eq8}
\Delta a_{\mu}=\frac{m^2_{\mu}}{m_{1}m_{2}}\bigg\{\frac{s_Lc_Rm_{2}[G(x_{11})-G(x_{21})]+s_Rc_Lm_{1}[G(x_{22})-G(x_{12})]}{s_Lc_Rm_{1}[H(x_{11})-H(x_{21})]+s_Rc_Lm_{2}[H(x_{22})-H(x_{12})]}\bigg\},
\end{equation}
where $x_{ij}= (\frac{m_{i \mu}}{m_{j}})^2$ and
\begin{equation}\label{eq9}
G(x)= \frac{2x\ln x}{(x-1)^3}-\frac{x+1}{(x-1)^2}.
\end{equation}
In the simplifying case we are considering,
Eq. (18) is independent of $\theta_L=\theta_R$. 
In Fig.~3 we plot $m_{1 \mu}$ against $m_1$ 
for various ratios $m_{2 \mu}/m_{1 \mu}$ 
in order to show the values of 
$m_1$ and $m_{1,2 \mu}$ 
which can account for 
the discrepancy between the 
experimental measurement~\cite{g-2exp}
and the SM prediction~\cite{g-2th}
\begin{equation}\label{eq20}
\Delta a_{\mu} = 39.35 \pm 5.21_{{\rm th}} \pm 6.3_{{\rm exp}} \times 10^{-10}
\end{equation}
We have combined the experimental and theoretical
uncertainties in quadrature, which corresponds to 
the curved limits of the shaded regions.
The lower limit of 200 GeV for $m_1$ corresponds to 
$\theta_L = \pi/4$.
\begin{figure}[htb] 
\centering
\includegraphics[scale=.6]{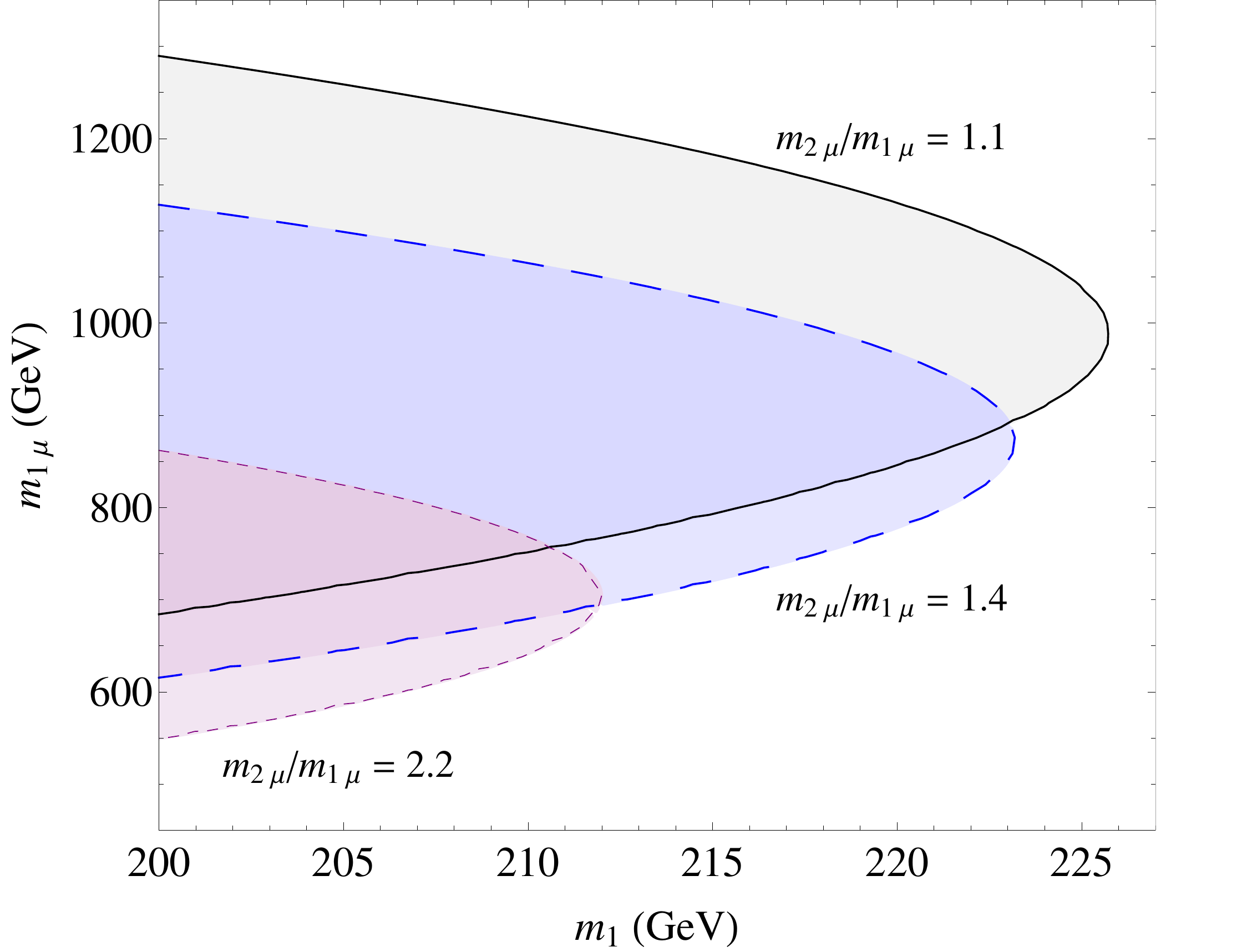}
\caption{Values of $m_1$ and $m_{1,2 \mu}$ which can explain $\Delta a_{\mu}$ for the case $\theta_L=\theta_R$.}
\end{figure}

The subdominant contributions to $\Delta a_{\mu}$ from $f'^2$ , ￼￼and $f_{\mu}^2$ are negative as expected , i.e.
\begin{eqnarray}\label{eq10}
(\Delta a_{\mu})'=\frac{-m^2_{\mu}}{32\pi^2}\bigg\{&f'^2\bigg[\frac{s^2_L}{m^2_1}\bigg(c^2_{\mu}J(x_{11})+s^2_{\mu}J(x_{21})\bigg)+\frac{c^2_L}{m^2_2}\bigg(c^2_{\mu}J(x_{12})+s^2_{\mu}J(x_{22})\bigg)\bigg]\nonumber\\&+ f^2_{\mu}\bigg[\frac{c^2_R}{m^2_1}\bigg(s^2_{\mu}J(x_{11})+c^2_{\mu}J(x_{21})\bigg)+\frac{s^2_R}{m^2_2}\bigg(s^2_{\mu}J(x_{12})+c^2_{\mu}J(x_{22})\bigg)\bigg]\bigg\},
\end{eqnarray}
where
\begin{equation}\label{eq11}
J(x)= \frac{x\ln x}{(x-1)^4}+\frac{x^2-5x-2}{6(x-1)^3}.
\end{equation}
The third contribution is from $s$ exchange which will be introduced in the next section and is given by
\begin{equation}\label{eq12}
(\Delta a_{\mu})''=\sum^3_{i=1} \frac{-f^2|U_{\mu i}|^2m^2_{\mu}}{16\pi^2m_{E}^2}G_{\gamma}(x_{i}),
\end{equation}
where $x_i=\frac{m^2_{s_i}}{m^2_{E}}$ and
\begin{equation}\label{eq13}
G_{\gamma}(x)=\frac{ 2 x^3+3 x^2-6 x^2 \ln x-6 x+1}{6(x-1)^4}<\frac{1}{6}.  
\end{equation}
The mass of $E^-$ has a lower limit of $m_E \simeq$ 300 GeV, 
which is numerically equivalent to 
$G_F m_E^2 \simeq 1$ used in the following section,
due to our parameterization for the fermion mixing
of $N$ and $E^0$.  
Hence $(\Delta a_{\mu})''$ is less than $10^{-10}f^2$, 
which for $f<1$ is below the present experimental sensitivity of $10^{-9}$ 
and thus can be neglected.

\section{Rare Lepton Decays}

Whereas $Z_3$ lepton triality is exact in Fig.~1, the corresponding diagram 
for neutrino mass breaks it, as shown below.
\begin{figure}[htb]
\vspace*{-3cm}
\hspace*{-3cm}
\includegraphics[scale=1.0]{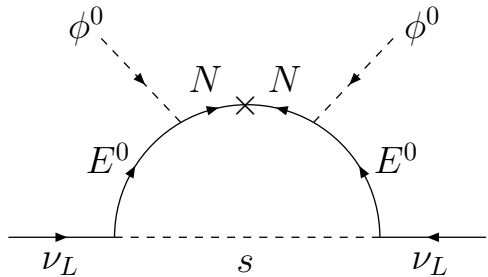}
\vspace*{-21.5cm}
\caption{One-loop generation of neutrino mass.}
\end{figure}
The new particles are three real scalars $s_{1,2,3} \sim \underline{3}$ 
under $A_4$. To connect the loop, Majorana mass terms $(m_L/2) N_L N_L$ and 
$(m_R/2) N_R N_R$ are assumed.  Since both $E$ and $N$ may be defined 
to carry lepton number, these new terms violate lepton number softly and 
may be naturally small. Using the Yukawa interaction 
$f s \bar{E}^0_R \nu_L$, the one-loop Majorana neutrino mass is given by 
\begin{eqnarray}
m_\nu &=& f^2 m_R \sin^2 \theta_R \cos^2 \theta_R (m_1^2 - m_2^2)^2 
\int {d^4 k \over (2 \pi)^4} {k^2 \over (k^2 - m_s^2)} {1 \over 
(k^2 - m_1^2)^2} {1 \over (k^2 - m_2^2)^2} \nonumber \\ 
&+& f^2 m_L m_1^2 \sin^2 \theta_L \cos^2 \theta_R \int {d^4 k \over (2 \pi)^4} 
{1 \over (k^2 - m_s^2)}{1 \over (k^2 - m_1^2)^2} \\ 
&+& f^2 m_L m_2^2 \sin^2 \theta_R \cos^2 \theta_L \int {d^4 k \over (2 \pi)^4}
{1 \over (k^2 - m_s^2)}{1 \over (k^2 - m_2^2)^2} \nonumber \\ 
&-& 2 f^2 m_L m_1 m_2 \sin \theta_L \sin \theta_R \cos \theta_L \cos \theta_R 
\int {d^4 k \over (2 \pi)^4}{1 \over (k^2 - m_s^2)}{1 \over (k^2 - m_1^2)}
{1 \over (k^2 - m_2^2)}. \nonumber  
\end{eqnarray}
This formula holds for $s$ as a mass eigenstate.  If $A_4$ is unbroken, then 
$s_{1,2,3}$ all have the same mass and ${\cal M}_\nu$ is proportional to the 
identity matrix.  However, if $A_4$ is softly broken by the necessarily 
real $s_i s_j$ mass terms, then the neutrino mass matrix is given by
\begin{equation}
{\cal M}_\nu = {\cal O} \pmatrix{m_{\nu1} & 0 & 0 \cr 0 & m_{\nu2} & 0 \cr 
0 & 0 & m_{\nu3}} {\cal O}^T,
\end{equation}
where ${\cal O}$ is an orthogonal matrix and ${\cal O} \neq 1$ breaks $Z_3$ 
lepton triality explicitly. 
Now each $m_{\nu i}$ may be complex because $f$, $m_L$, $m_R$ may be 
complex, but a common unphysical phase, say for $\nu_1$, 
may be rotated away, leaving just two relative Majorana phases for 
$\nu_2$ and $\nu_3$, owing to the relative phase between $m_L$ and 
$m_R$ with different $s_{1,2,3}$ masses.  Hence ${\cal M}_\nu$ 
is diagonalized by ${\cal O}$, which is all that is required to obtain 
\underline{cobimaximal} mixing~\cite{m15-2}, i.e. 
$\theta_{23} = \pi/4$ and $\delta_{CP} = \pm \pi/2$, once $U_\omega$ is 
applied, as explained in Ref.~\cite{m15-1}.

The companion interaction to $f s \bar{E}_R^0 \nu_L$ is $f s \bar{E}_R^- l_L$, 
which induces the radiative process $l_i \to l_j + \gamma$.  In the limit 
of exact $Z_3$ lepton triality, this amplitude is zero.  Here it is 
proportional to $\sum_k U_{ik} U^*_{jk} F_k$ where $F_{1,2,3}$ refer to 
functions of $m^2_{s_{1,2,3}}$, and $U_{ik}$ is the neutrino mixing matrix. 
Clearly, it is also zero if $F_1 = F_2 = F_3$. The amplitude for $\mu \to e \gamma$ is given by
\begin{equation}\label{eq14}
A_{\mu e}=\frac{ef^2m_{\mu}}{32\pi^2m_{E}^2}\sum_iU^*_{ei}U_{\mu i}G_{\gamma}(x_i),
\end{equation}
Using the most recent $\mu \to e \gamma$ bound~\cite{MEG13}, this branching 
fraction is constrained by  
\begin{equation}\label{eq15}
B=\frac{12\pi^2|A_{\mu e}|^2}{m^2_{\mu}G^2_F}<5.7\times10^{-13}.
\end{equation}
For small $x_i$ and $x_1 \simeq x_2$, 
\begin{equation}
|\sum_i U^*_{ei} U_{\mu i} G_\gamma (x_i)| = {s_{13} c_{13} \over 3 \sqrt{2}} 
|x_3 - x_2|,
\end{equation}
where $s_{13} = \sin \theta_{13}$, $c_{13} = \cos \theta_{13}$, and 
$\sin \theta_{23} = 1/\sqrt{2}$ has been assumed.  Hence
\begin{equation}
B = {\alpha s_{13}^2 c_{13}^2 \over 384 \pi} \left( {f^2 |x_3 - x_2| \over 
G_F m_E^2} \right)^2.
\end{equation}
Let $G_F m_E^2 \simeq 1$, $f = 0.2$, $|x_3 - x_2| \simeq 0.05$, 
then $B = 5.6 \times 10^{-13}$, just below the experimental constraint.

Another possible rare decay is $\mu \to eee$, which comes from 
$\mu \to e (\gamma , Z) \to eee$ as well as directly through a box diagram 
as shown below. \begin{figure}[htb]
\vspace*{-3cm}
\hspace*{-3cm}
\includegraphics[scale=1.0]{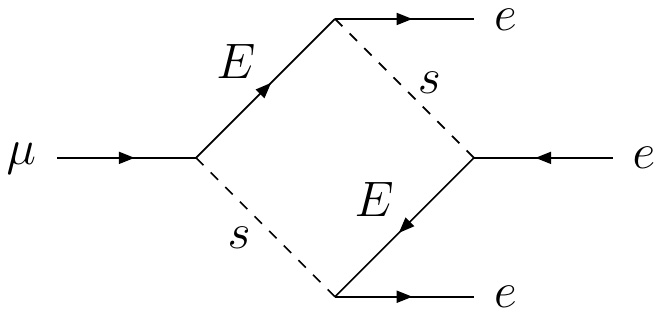}
\vspace*{-21.5cm}
\caption{Box diagram for $\mu \to eee$.}
\end{figure}
The amplitude for the former process with a virtual photon is given by
\begin{eqnarray}\label{eq17}
i\mathcal{M}_{\gamma} &=& \frac{-ie^2f^2}{32\pi^2m_{E}^2}\sum_{i=1}^3U^*_{ei}U_{\mu i}\bar{u}(p_1)\left[G_e(x_i)\left(\gamma^{\alpha}-\frac{q^{\alpha}\slashed{q}}{q^2}\right)P_L-im_{\mu}G_\gamma(x_i)\frac{\sigma^{\alpha\beta}q_{\beta}}{q^2}P_R\right]u_{\mu}(p)\bar{u}(p_2)\gamma_{\alpha}v(p_3)\nonumber\\ &-& (p_1 \leftrightarrow p_2),
\end{eqnarray}
where $P_{L,R} = (1 \mp \gamma_5)/2$, $q=p-p_1$ and 
\begin{equation}\label{eq18}
G_e(x)= \frac{7-36 x+45 x^2-16 x^3+6 x^2(2 x-3)\ln x}{18(x-1)^4}.
\end{equation}
The amplitude for the process with a virtual $Z$ boson has a similar form 
because $E_{L,R}$ is vector-like, but it is further suppressed by $m_Z^2$. 
The amplitude for the box diagram is given by
\begin{equation}\label{eq19}
i\mathcal{M}_B=\frac{if^4[\bar{u}(p_1)\gamma_{\alpha}P_Lu_{\mu}(p)\bar{u}(p_2)\gamma^{\alpha}P_Lv(p_3)-(p_1 \leftrightarrow p_2)]}{64\pi^2m^2_E}\sum_{i,j=1}^3U_{\mu i}U^*_{ej}[U_{ei}U^*_{ej}-U_{ej}U^*_{ei}]B_{ij},~
\end{equation}
where
\begin{equation}\label{eq20}
B_{ij}= \frac{B(x_i)-B(x_j)}{x_i-x_j} ~~ i\neq j, ~~
B_{ii} = \frac{x_i^2-2x_i\ln x_i-1}{(x_i-1)^3},~~ B(x) = \frac{x^2\ln x}{(x-1)^2}-\frac{1}{x-1}.
\end{equation}

With the same specific choice of parameters as in Eq.~(29) we find that the box diagram contribution is dominant. Hence
the $\mu \to eee$ branching fraction is
\begin{equation}\label{eq21}
B'=\frac{f^8}{2(8\pi)^4 m^4_EG^2_F} \bigg|\sum_{i,j=1}^3U_{\mu i}U^*_{ej}[U_{ei}U^*_{ej}-U_{ej}U^*_{ei}]B_{ij}\bigg|^2.~
\end{equation}
Using the bound~\cite{SINDRUM} on $\mu \to e e e$ decay and for small $x_i$ we have
 \begin{equation}\label{eq22}
B'=\frac{f^8}{2(8\pi)^4 m^4_EG^2_F} \frac{\sin^2(4\theta_{13})}{8}<1.0\times 10^{-12}.~
\end{equation}
This constraint is easily satisfied for $G_F m_E^2 \simeq 1$, $f = 0.2$, 
which yields $B' = 1.35 \times 10^{-13}$.

\section{Dark Matter}

As for dark matter, there is a one-to-one correlation of the neutrino mass 
eigenstates to the $s_{1,2,3}$ mass eigenstates, the lightest of which is 
dark matter.  Due to the presence of the $A_4$ symmetry, the dark matter 
parity of this model is also derivable from lepton parity~\cite{m15-3}.  
Under lepton parity, let the new particles $(E^0,E^-), N$ be even and 
$s,x,y$ be odd, then the same Lagrangian is obtained.  As a result, 
dark parity is simply given by $(-1)^{L+2j}$, which is odd for all the 
new particles and even for all the SM particles.  Note that the tree-level 
Yukawa coupling $\bar{l}_L l_R \phi^0$ would be allowed by lepton parity alone, 
but is forbidden here because of the $A_4$ symmetry.

If the Yukawa coupling $f$ of $s$ to leptons is small, its relic density 
and elastic cross section off nuclei are both controlled by the interaction 
$\lambda v h s^2$.  As such, a recent analysis~\cite{fpu15} claims that the 
resulting allowed parameter space is limited to a small region near 
$m_s < m_h/2$.  To evade this constraint, the mechanism of Ref.~\cite{m15-4} 
may be invoked.  Add a complex neutral singlet scalar $\chi \sim 
\underline{1}'$ under $A_4$ with $Z_2$ even.  The dimension-four terms 
of the Lagrangian are of course required to be invariant under $A_4$. 
We assume that the dimension-three terms are also invariant: $\chi^3$, 
$(\chi^\dagger)^3$, $(s_1^2 + \omega^2 s_2^2 + \omega s_3^2)\chi$, and 
$(s_1^2 + \omega s_2^2 + \omega^2 s_3^2)\chi^\dagger$.  The symmetry $A_4$ 
is broken only by the dimension-two terms: $\chi^2$, $(\chi^\dagger)^2$, 
and $s_i s_j$.  As a result, $\chi$ is split into $\chi_R$ and $\chi_I$, 
each mixing with $h$ radiatively.  In the physical basis, the dark matter $s$ 
has residual $s^2 \chi_{R,I}$ interactions which contribute to its 
annihilation cross section, but do not affect its scattering off nuclei 
through $h$ exchange. 

Let us denote the $\chi_{R,I}$ masses with $m_{R,I}$. For illustration, we assume $m_{R} < m_s < m_{I}$, and take the $\chi_I\chi_R^2$ coupling to be zero, so that the annihilations shown in Fig. 6 are controlled by the interaction terms
\begin{equation}
-\mathcal{L}_{int} = \frac{\lambda'}{4}\, s^2\chi^2_R +\, \frac{g}{2} s^2\chi_R +\, \frac{g'}{3!}\chi_R^3
\end{equation}
\begin{figure}[htb]
\includegraphics[scale=0.84]{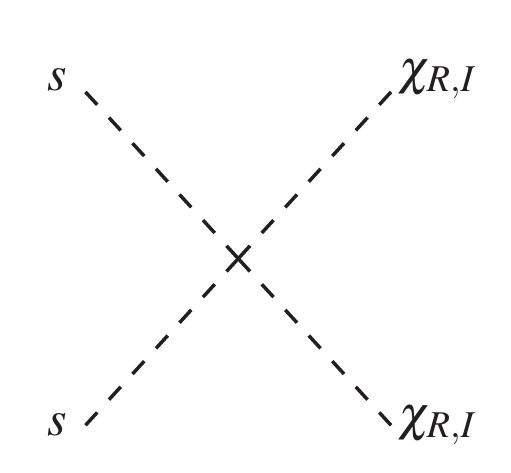}
\hfill
\includegraphics[scale=1.0]{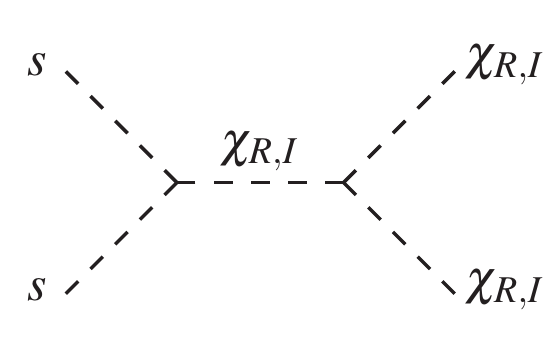}
\hfill
\includegraphics[scale=1.0]{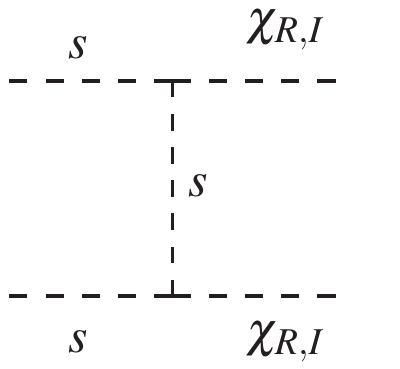}
\caption{$s\,s$ annihilation to $\chi_{R,I}$ mass eigenstates.}
\end{figure}

As a result, the annihilation cross section times relative velocity is given by
\begin{equation}
\sigma\times v_{rel} =\, \frac{\sqrt{1-(m_{R}/m_s)^2}}{64\pi m_s^2}\left(\lambda' + \frac{g' g}{4 m_s^2 - m^2_{R}} - \frac{g^2}{2 m_s^2 - m^2_{R}}\right)^2.
\end{equation}
Setting this equal to $2.2 \times 10^{-26}\, \mbox{cm}^3 \mbox{s}^{-1}$, with $m_s = 200$ GeV and $m_{R} = 150 $ GeV, we find 
\begin{equation}
\lambda' + 0.073 \left(\frac{\sqrt{g' g}}{100\, \mbox{GeV}}\right)^2 - 0.174 \left(\frac{g}{100\, \mbox{GeV}}\right)^2 = 0.1514.
\end{equation}
Note that $\chi_R$ decays to SM particles through its mixing with $h$. As mentioned earlier, the spin-independent elastic cross section proceeds through $h$ exchange, with 
\begin{equation}
\sigma_{SI} = \frac{\lambda^2 f_N^2 \mu^2 m_N^2}{\pi m_h^4 m_s^2},
\end{equation}
where $\mu = m_N m_s / (m_N + m_s)$ is the DM-nucleon reduced mass,  $m_N = (m_p + m_n)/2 = 938.95 \, \mbox{MeV}$ is the nucleon mass, and $f_N = 0.3$ is the
Higgs-nucleon coupling factor~\cite{x}. The LUX bound~\cite{LUX15} for $m_s = 200 \, \mbox{GeV}$ is $\sigma \approx 1.5$ zb , which implies
\begin{equation}
\lambda <  3.3\times 10^{-4}.
\end{equation}

In conclusion, in the context of a specific $A_4$ scotogenic (dark-matter-induced) model of 
radiative neutrino and charged-lepton masses with the one Higgs boson of 
the standard model, we study finite calculable anomalous Higgs couplings 
with possible large deviations from the SM predictions.  We show that the 
observed discrepancy in the muon anomalous magnetic moment may be explained 
by new particles in the TeV mass range, with predictions for the lepton 
flavor violating processes $\mu \to e \gamma$ and $\mu \to eee$.  We also 
discuss the nature of the expected dark matter in this scenario.

\medskip

This work is supported in part by the U.~S.~Department of Energy under 
Grant No.~DE-SC0008541.
\newpage
\baselineskip 18pt
\bibliographystyle{unsrt}

\end{document}